\documentclass[runningheads]{llncs}
\usepackage{graphicx}
\usepackage[dvipsnames]{xcolor}
\usepackage{bm}
\usepackage{mathtools}
\usepackage{amsfonts}
\usepackage{enumitem}
\usepackage{xspace}

\usepackage{filecontents}
\usepackage[dvipsnames]{xcolor}
\usepackage{hyperref}
\usepackage{cleveref}

\newcommand\myshade{85}
\colorlet{mylinkcolor}{BrickRed}
\colorlet{mycitecolor}{NavyBlue}
\colorlet{myurlcolor}{Aquamarine}

\hypersetup{
  linkcolor  = mylinkcolor!\myshade!black,
  citecolor  = mycitecolor!\myshade!black,
  urlcolor   = myurlcolor!\myshade!black,
  colorlinks = true,
}

\newcommand{\past}[2]{\cev{\bm #1}_{#2}}

\newcommand\independent{\protect\mathpalette{\protect\independenT}{\perp}}
\def\independenT#1#2{\mathrel{\rlap{$#1#2$}\mkern2mu{#1#2}}}

\DeclareMathOperator*{\argmin}{arg\,min}

\makeatletter
\DeclareRobustCommand{\cev}[1]{%
  \mathpalette\do@cev{#1}%
}
\newcommand{\do@cev}[2]{%
  \fix@cev{#1}{+}%
  \reflectbox{$\m@th#1\vec{\reflectbox{$\fix@cev{#1}{-}\m@th#1#2\fix@cev{#1}{+}$}}$}%
  \fix@cev{#1}{-}%
}
\newcommand{\fix@cev}[2]{%
  \ifx#1\displaystyle
    \mkern#23mu
  \else
    \ifx#1\textstyle
      \mkern#23mu
    \else
      \ifx#1\scriptstyle
        \mkern#22mu
      \else
        \mkern#22mu
      \fi
    \fi
  \fi
}
\makeatother

\newcommand{\Ex}{\mathbb{E}}

\newcommand{\MIfull}{Causal blanket}
\newcommand{\mifull}{causal blanket}
\newcommand{\MIfulls}{{\MIfull}s}
\newcommand{\MI}{CB\xspace}
\newcommand{\MIs}{{\MI}s\xspace}

\newcommand\figsubref[2]{\hyperref[#1]{\ref*{#1}#2}}

\begin{document}
\title{\MIfulls:
\\Theory and algorithmic framework}

\author{Anonymous authors\\}

\author{Fernando E. Rosas\inst{1,2,3} \and
Pedro A.M. Mediano\inst{4} \and
Martin Biehl\inst{5} \and\\
Shamil Chandaria\inst{1,6}
\and
Daniel Polani\inst{7}
\thanks{
\scriptsize{F.R. was supported by the Ad Astra
Chandaria foundation. P.M. was funded by the Wellcome Trust (grant no.
210920/Z/18/Z). M.B. was supported by a grant from Templeton World Charity
Foundation, Inc. (TWCF). The opinions expressed in this publication are those
of the authors and do not necessarily reflect the views of TWCF. }}}
\authorrunning{Rosas, Mediano, Biehl, Chandaria, and Polani}

\institute{Institutions}
\institute{Centre for Psychedelic Research, Imperial College London, London SW7 2DD, UK \and
Data Science Institute, Imperial College London, London SW7 2AZ, UK \and
Centre for Complexity Science, Imperial College London, London SW7 2AZ, UK
\and
Department of Psychology, University of Cambridge, Cambridge CB2 3EB, UK\\
\and
Araya Inc., Tokyo 107-6024, Japan\\
\and
Institute of Philosophy, School of Advanced Study, University of London, UK\\
\and
Dept. of Computer Science, University of Hertfordshire, Hatfield AL10 9AB, UK\\
\email{f.rosas@imperial.ac.uk ~ pam83@cam.ac.uk ~ martin@araya.org shamil.chandaria@gmail.com ~ d.polani@herts.ac.uk}
}
\maketitle

\begin{abstract}

We introduce a novel framework to identify perception-action loops (PALOs)
directly from data based on the principles of computational mechanics. Our
approach is based on the notion of \textit{causal blanket}, which captures
sensory and active variables as dynamical sufficient statistics --- i.e.\ as
the ``differences that make a difference.'' Furthermore, our theory provides a
broadly applicable procedure to construct PALOs that requires neither a
steady-state nor Markovian dynamics. Using our theory, we show that every
bipartite stochastic process has a causal blanket, but the extent to which this
leads to an effective PALO formulation varies depending on the integrated
information of the bipartition.

\end{abstract}

\section{Introduction}\label{sec:1}

The perception-action loop (PALO) is one of the most important constructs of
cognitive science, and plays a fundamental role in many other disciplines
including reinforcement learning and computational neuroscience. Despite its
importance and pervasiveness, fundamental questions about what kind of systems
can be properly described by a PALO are still to a large extent unanswered. The
aim of this paper is to introduce a framework that allows us to identify PALOs
directly from data, which complements existent approaches and serves to deepen
our understanding of the essential elements that make a PALO.

\subsection{Markov blankets}
\label{sec:1a}

One of the most encompassing accounts of PALOs can be found in the Free Energy
Principle (FEP) literature, which formalises them via \emph{Markov blankets}
(MBs)~\cite{kirchhoff2018markov}. An interesting contribution of this
literature is to characterise ``sensory'' ($S$) and ``active'' ($A$) variables
as having two defining properties: (i) they mediate the interactions between
internal variables of the agent ($M$) and external variables of its environment
($E$), and (ii) they impose a specific causal structure on these interactions
--- e.g. sensory variables may affect internal variables, but are not
(directly) affected by them~\cite{kirchhoff2018markov}.

Formally, MBs were originally introduced by Pearl~\cite{pearlprobabilistic1988}
for Markov and Bayesian networks. Within the FEP literature, MBs are usually
employed in multivariate stochastic processes with ergodic Markovian dynamics,
with a steady-state distribution $p^*$ that is required to
satisfy~\cite{parrmarkov2020}
\begin{align}
\label{eq:mbfactorization}
  p^*(e_t,m_t|s_t,a_t)=p^*(e_t|s_t,a_t)p^*(m_t|s_t,a_t)~.
\end{align}
However, Eq.~\eqref{eq:mbfactorization} does not suffice to guarantee a PALO
structure, as noted in Ref.~\cite{biehltechnical2020}. In effect, the MB
condition is insufficient to establish requirement (ii): its symmetry with
respect to internal and external variables makes it impossible to infer the
direction of the loop; additionally, the fact that the condition holds across
variables synchronously makes it unsuitable to guarantee a causal
relationship~\cite{pearl2009causality}. Recent reports~\cite{friston2020some}
acknowledge that this synchronous condition needs to be complemented with
additional diachronic restrictions on the system's dynamics, which can be
written, for instance, as a set of coupled stochastic differential equations of
the form
\begin{align}
\label{eq:mbvectorfield}
\begin{split}
    \dot{m}_t = f_{\text{in}}(m_t,a_t,s_t) + \omega_t^{\text{in}}~,~~~~~
    &\dot{a}_t = f_{\text{a}}(m_t,a_t,s_t) + \omega_t^{\text{a}}~,\\
    \dot{e}_t = f_{\text{ex}}(e_t,a_t,s_t) + \omega_t^{\text{ex}}~,~~~~~
    &\dot{s}_t = f_{\text{s}}(e_t,a_t,s_t) + \omega_t^{\text{s}}~.
    \end{split}
\end{align}
Above, the functions $f_{\text{in}},f_{\text{a}},f_{\text{ex}},f_{\text{s}}$
determine the flow, and
$\omega_t^{\text{in}},\omega_t^{\text{a}},\omega_t^{\text{ex}},\omega_t^{\text{s}}$
denote additive Gaussian noise. Interestingly, it has been shown that
Eq.~\eqref{eq:mbvectorfield} implies Eq.~\eqref{eq:mbfactorization} under
additional assumptions: either block diagonality conditions over the solenoidal
flow~\cite{friston2020some}, or strong
dissipation~\cite[Appendix]{friston2020parcels}.\footnote{However, in the
general case neither Eqs.~\eqref{eq:mbfactorization} or
\eqref{eq:mbvectorfield} imply each other~\cite{biehltechnical2020} --- hence
they need to be taken as complementary conditions.} Hence, PALOs could be
interpreted as coupled stochastic dynamical systems of the form in
Eq.~\eqref{eq:mbvectorfield}, as long as the flow satisfies any of the two
mentioned conditions.

\begin{figure}[t!]
  \centering
  \includegraphics{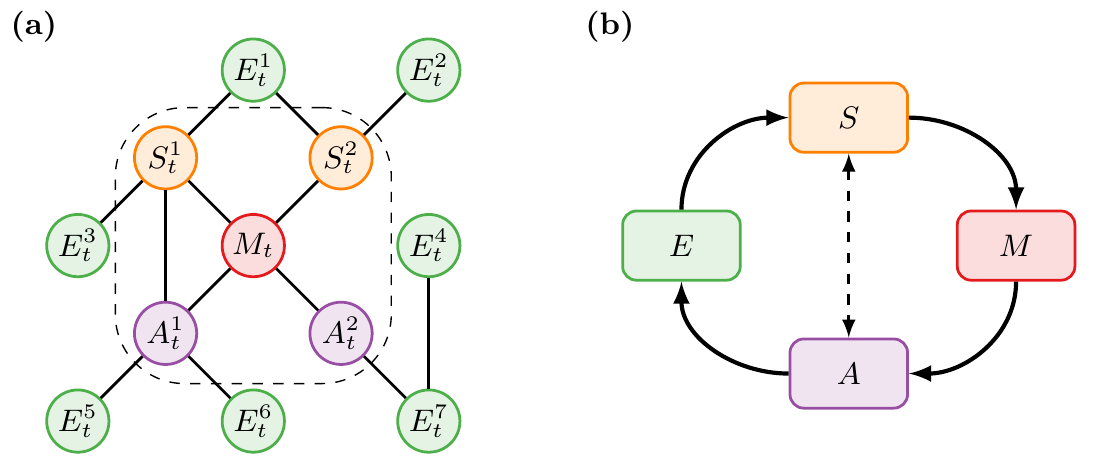}
  \vspace{-5pt}
  \caption{
  Two visualisations of PALOs in the FEP literature, either based on 
  \textbf{(a)} Markov blankets according to Eq.~\eqref{eq:mbfactorization} 
  or \textbf{(b)} Langevin dynamics following Eq.~\eqref{eq:mbvectorfield}.
  }
  \label{fig:badblanket}
\end{figure}

Despite its elegance, this formalisation of PALOs has important limitations.
First, this formulation relies strongly on Langevin dynamics, making it
difficult to extend it to PALOs appearing in discrete systems. Secondly, this
approach depends on a set of assumptions --- for one, the aforementioned
conditions over the flow and the restriction to systems in their steady-state
--- that might be too restrictive for some scenarios of interest. Finally, and
perhaps most importantly, Eq.~\eqref{eq:mbfactorization} forces all
interactions between $M_t$ and $E_t$ to be accountable by $(S_t,A_t)$, which
imposes --- due to the data processing inequality~\cite{cover2012elements} ---
an information bottleneck of the form $I(M_t;E_t) \leq I(M_t; A_t,S_t)$.
Therefore, the MB formalism forbids interdependencies induced by past events
that are kept in memory, but may not directly influence the present state of
the blankets.\footnote{We thank Nathaniel Virgo for first noting this issue.}
This information kept in memory arguably plays an important role in many PALOs,
and includes uncontroversial features of cognition (such as old memories that
an agent retains but is neither caused by a sensation nor causing an action at
the current moment), yet are forbidden by MBs.

\subsection{Computational mechanics, causal states, and epsilon-machines}
\label{sec:1b}

Computational mechanics is a method for studying patterns and statistical
regularities observed in stochastic processes by uncovering their hidden causal
structure~\cite{shalizi2001computational,shalizi2001causal}. A key insight is
that an optimal, minimimal representation of a process can be revealed by
grouping past trajectories according to their forecasting abilities into
so-called \textit{causal states}. More precisely, the causal states of a
(possibly non-Markovian) time series $\{Z_t\}_{t\in\mathbb{Z}}$ are the
equivalent classes of trajectories $\past{z}{t} \coloneqq (\dots, z_{t-1},
z_{t})$ given by the relationship
\begin{equation}
    \past{z}{t} \equiv_\epsilon \past{z}{t}' \qquad \text{iff} \qquad p(z_{t+1} | \past{z}{t}) = p(z_{t+1} | \past{z}{t}') \quad \forall z_{t+1} ~.
\nonumber
\end{equation}
It can be shown that the causal states are the coarsest coarse-graining of past
trajectories $\past{x}{t}$ that retains full predictive power over future
variables~\cite{crutchfield1989inferring,grassberger1986toward}. Moreover, the
corresponding process over causal states always has Markovian dynamics,
providing the simplest yet encompassing representation of the system's
information dynamics on a latent space --- known as the
\textit{epsilon-machine}.

Please note that the causal states of a system are guaranteed to provide
counterfactual relationships~\cite{pearl2009causality} only if the system at
hand is fully observed. In the case of partially observed scenarios, causal
states ought to be understood in the Granger sense, i.e. as states of maximal
non-mediated predictive ability~\cite{bressler2011wiener}.

\subsection{Contribution}

In this paper we introduce an operationalisation of PALOs based on \emph{causal
blankets} (\MI), a construction based on a novel definition of dynamical
statistical sufficiency. CB capture properties (i) and (ii) in a single
mathematical construction by applying informational constructs directly to
dynamical conditions. Moreover, CBs can be constructed with great generality
for any bipartite system without imposing further conditions, and hence can be
applied to non-ergodic, non-Markovian stochastic processes. This generality
allows us to explore novel connections between PALOs and integrated
information. In the rest of the manuscript, we:
\begin{enumerate}

  \item[1)] Provide a rigorous definition of \MIs (Definition~\ref{def:MB}); and

  \item[2)] Show every agent-environment partition has a \MI, and thus can be
  described as a PALO (Proposition~\ref{prop:PALO}); although

  \item[3)] Not all systems are equally well described as a PALO, and this can
  be quantified via information geometry and integrated information
  (Sec.~\ref{sec:BeyondBlankets}) --- providing a principled measure to
  distinguish preferable candidates for PALO.\footnote{The proofs of our
  results can be found in the Appendix.} 

\end{enumerate}

\section{Causal blankets as informational boundaries}

We consider the perspective of a scientist who repeatedly measures a system
composed of two interacting parts $X_t$ and $Y_t$. We assume that, from these
observations, a reliable statistical model of the corresponding discrete-time
stochastic process can be built --- of which all the resulting marginal and
conditional distributions are well-defined. Random variables are denoted by
capital letters (e.g. $X,Y$) and their realisations by lower case letters (e.g.
$x,y$); stochastic processes at discrete times (i.e.\ time series) are
represented as bold letters without subscript $\bm X =
\{X_t\}_{t\in\mathbb{Z}}$, and $\past{X}{t} \coloneqq (\dots, X_{t-1}, X_{t})$
denotes the infinite past of $\bm X$ until and including $t$.

Given two random variables $X$ and $Y$, a statistic $U=f(X)$ is said to be
\emph{Bayesian sufficient of $X$ w.r.t. $Y$} if $X \independent Y \mid U$,
which implies that all the common variability between $X$ and $Y$ is accounted
for by $U$~\cite{cover2012elements}. The first step in our construction is to
introduce a dynamical version of statistical sufficiency.

\begin{definition}[D-BaSS]\label{def:DBass}
Given two stochastic processes $\bm X,\bm Y$, a process $\bm U$ is a dynamical
Bayesian sufficient statistic (D-BaSS) of $\bm X$ w.r.t. $\bm Y$ if,
for all $t\in\mathbb{Z}$, the following conditions hold:
\begin{enumerate}[label=\roman*.]

\item Precedence: there exists a function $F(\cdot)$ such that $U_t =
F(\past{X}{t})$ for all $t\in\mathbb{Z}$.

\item Sufficiency: $Y_{t+1} \independent \past{X}{t} \mid (U_{t},
\past{Y}{t} )$~.

\end{enumerate}
Moreover, a stochastic process $\bm M$ is a minimal D-BaSS of $\bm X$ with
respect to $\bm Y$ if it is itself a D-BaSS and for any D-BaSS $\bm U$ there
exists a function $f(\cdot)$ such that $f(U_t) = M_t, \forall t\in\mathbb{Z}$.
\end{definition}

The first condition above states that $\bm U$ is no more than a simpler,
coarse-grained representation of $\bm X$, and the second implies that the
influence of $\past{X}{t}$ on $Y_{t+1}$ given $\past{Y}{t}$ is fully mediated
by $U_t$. This has interesting consequences for transfer entropy, as seen in
the next lemma.
\begin{lemma}\label{lemma:te}
If $\bm U$ is a D-BaSS of $\bm X$ w.r.t. $\bm Y$, then
\begin{equation}
\text{\textnormal{TE}}(\bm X\rightarrow \bm Y)_t\coloneqq I(\past{X}{t} ; Y_{t+1} | \past{Y}{t} ) = I(U_{t} ; Y_{t+1} | \past{Y}{t} )~.
\end{equation}
\end{lemma}

There are many such D-BaSS; e.g.\ $U_t = \past{X}{t}$ would be one valid D-BaSS
of $\bm X$ w.r.t. $\bm Y$. However, Theorem~\ref{prop:unique} shows that
minimal D-BaSS's are unique (up to bijective transformations).

\begin{theorem}[Existence and uniqueness of the minimal D-BaSS]\label{prop:unique}
Given stochastic processes $\bm X,\bm Y$, the minimal
D-BaSS of $\bm X$ w.r.t. $\bm Y$ corresponds to the partition
of past-trajectories $\past{x}{t}$ induced by the following equivalence
relationship:
\begin{equation}
\past{x}{t} \equiv_p \past{x}{t}' \qquad \text{iff} \qquad \forall \past{y}{t}, y_{t+1} \quad  p(y_{t+1} | \past{x}{t}, \past{y}{t}) = p(y_{t+1} | \past{x}{t}', \past{y}{t})~.
\nonumber
\end{equation}
Therefore, the minimal D-BaSS is always well-defined, and is unique up to
an isomorphism.
\end{theorem}

This result shows that D-BaSSs can be built irrespective of any other possibly
latent influences on $\bm X$ and $\bm Y$, as it is defined purely on the joint
statistics of these two processes. Moreover, Theorem~\ref{prop:unique} provides
a recipe to build a D-BaSS: group together all the past trajectories that lead
to the same predictions, which is a key principle of computational
mechanics~\cite{crutchfield1989inferring,grassberger1986toward,shalizi2001computational,shalizi2001causal}.
Therefore, a minimal D-BaSS distinguishes only ``differences that make a
difference'' for the future dynamics, generalising the construction presented
in Ref.~\cite[Definition~1]{biehlaction2017} for Markovian dynamical systems,
and being closely related to the notion of sensory equivalence presented in
Ref.~\cite{ay2015umwelt}. With these ideas at hand, we can formulate our
definition of causal blanket.

\begin{definition}[\MIfull]\label{def:MB}
Given two stochastic processes $\bm X,\bm Y$, a \emph{reciprocal D-BaSS}
(ReD-BaSS) is a stochastic process $\bm R$ which satisfies:
\begin{enumerate}[label=\roman*.]

\item Joint precedence: $R_t = F(\past{X}{t}, \past{Y}{t})$ for some
function $F(\cdot)$.

\item Reciprocal sufficiency: $\bm R$ is a D-BaSS of $\bm X$ w.r.t. $\bm Y$, and also is a D-BaSS of $\bm Y$ w.r.t. $\bm X$.

\end{enumerate}
A \emph{\mifull} (\MI) is a minimal ReD-BaSS: a time series $\bm
M$, itself a ReD-BaSS, such that for all ReD-BaSSs $\bm R$ there exists a function $f(\cdot)$
such that $M_t = f(R_t), \forall t \in \mathbb{Z}$.
\end{definition}

This definition satisfies the two key desiderata discussed in
Section~\ref{sec:1a}: (i) a \MI mediates the interactions that take place
between $\bm X$ and $\bm Y$, and (ii) it assesses causality by focusing on
statistical relationships between past and future. From this perspective, \MIs
are the ``informational layer'' that causally decouples the agent's and
environment's temporal evolution from each other (see
Proposition~\ref{prop:trajectory}). Additionally, our next result guarantees
that \MIs always exist, and are unique to each bipartite system.

\begin{proposition}\label{prop:PALO}
Given $\bm X,\bm Y$, their \MI always exists and is unique (up to an
isomorphism). Moreover, their \MI is isomorphic to a pair $\{\bm S, \bm A\}$,
where $\bm A$ is a minimal D-BaSS of $\bm X$ w.r.t. $\bm Y$, and $\bm S$ is a
minimal D-BaSS of $\bm Y$ w.r.t. $\bm X$.
\end{proposition}

\begin{figure}[t]
\centering
\includegraphics{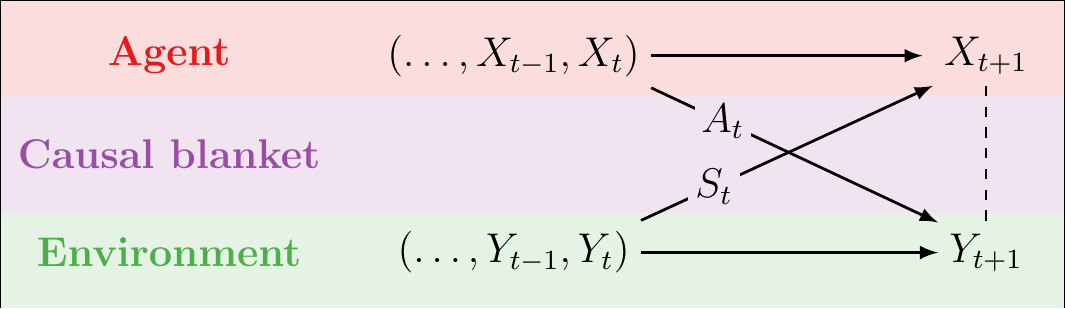}
\caption{Causal blanket $\{\bm S,\bm A\}$, which acts as a sufficient statistic mediating the interactions between $\bm X$ and $\bm Y$.}
\label{fig:PALO}
\end{figure}

Proposition~\ref{prop:PALO} has two important consequences: it guarantees that
\MIs \emph{always} exist, and that they naturally resemble a PALO --- as
visualised in Fig~\ref{fig:PALO}. Please note that this type of PALO
formalisation has a rich history, being studied in
Refs.~\cite{bertschinger2006information,bertschinger2008autonomy} and
variations being considered in
Refs.~\cite{klyubin2004organization,klyubin2007representations,tishby2011information}.
In contrast, our framework follows Refs.~\cite{ay2015umwelt,biehlaction2017}
and does not assume active and sensory variables as given, but discovers them
directly from the data. As a matter of fact, the ``sensory'' ($\bm S$) and
``active'' ($\bm A$) variables of CBs correspond (due to
Definition~\ref{def:MB}) to minimal sufficient statistics that mediate the
interdependencies between the past and future of $\bm X$ and $\bm Y$. The
construction of CBs imposes no requirements on the system's statistics or its
structure --- beyond the bipartition, holding also for non-ergodic and also
non-stationary systems, and systems with non-Markovian dynamics.

It is also possible to build internal and external states $M_t, E_t$ such that
$(M_t,A_t) = X_t$ and $(E_t,S_t)=Y_t$ with great generality. This can be done
via an orthogonal completion of the phase space; the details of this procedure
will be made explicit in a future publication. In this way, CBs can be thought
as suggesting implicit ``equations of motion'' somehow equivalent to
Eq.~\eqref{eq:mbvectorfield}, as shown in Figure~\ref{fig:PALO}. However, it is
important to remark that this representation does \emph{not} provide
counterfactual guarantees for partially observed systems (see
Section~\ref{sec:1b}).

\begin{example}
\label{ex:1}
Consider a multivariate stochastic process $\bm M,\bm A,\bm E,\bm S$ whose
dynamics follows
\begin{align}
\label{eq:discrete_mbvectorfield}
\begin{split}
    M_{t+1} = f_{\text{in}}(M_t,A_t,S_t) + N_{\text{in}},~~~~~
    & A_{t+1} = f_{\text{a}}(M_t,A_t,S_t) + N_{\text{a}},\\
    E_{t+1} = f_{\text{ex}}(E_t,A_t,S_t) + N_{\text{ex}},~~~~~
    &S_{t+1} = f_{\text{s}}(E_t,A_t,S_t) + N_{\text{s}},
    \end{split}
\end{align}
with $N_t^\text{in},N_t^\text{a},N_t^\text{ex},N_t^\text{s}$ being independent
of $M_t,A_t,E_t,S_t$ (note that Eq.~\ref{eq:discrete_mbvectorfield} corresponds
to a discrete-time version of Eq.~\eqref{eq:mbvectorfield}). Then, by defining
$X_t=(M_t,A_t)$ and $Y_t=(E_t,S_t)$, one can show using Definition~\ref{def:MB}
that that $\{\bm S,\bm A\}$ is the \MI of $\bm X,\bm Y$ --- as long as the
partial derivatives of $f_{\text{in}}, f_{\text{a}}, f_{\text{ex}},
f_{\text{s}}$ with respect to their corresponding arguments are nonzero.

\end{example}

\section{Integrated information transcends the blankets}
\label{sec:BeyondBlankets}

According to Def.~\ref{def:MB}, \MIs don't depend on the joint distribution
$p(x_{t+1},y_{t+1} | \past{x}{t}, \past{y}{t})$, but only on the marginals
$p(x_{t+1} | \past{x}{t}, \past{y}{t})$ and $p(y_{t+1} | \past{x}{t},
\past{y}{t})$. Here we study how meaningful the \MI (and the description of the
system as a PALO) is when the joint process's dynamics are different from the
product of these two marginals.

Let us start by introducing the \emph{synergistic coefficient} $\xi_t\in
\mathbb{R}$, which is a random variable given by
\begin{equation}\label{def:syn}
\xi_t \coloneqq \: \log \frac{ p(X_{t+1},Y_{t+1} | \past{X}{t}, \past{Y}{t}) }{ p(X_{t+1} | \past{X}{t}, \past{Y}{t}) \: p(Y_{t+1} | \past{X}{t}, \past{Y}{t}) }~.
\end{equation}
A process $(\bm X,\bm Y)$ is said to have \emph{factorisable dynamics} if
$\xi_t = 0$ \textit{a.s.} for all $t\in \mathbb{Z}$.

\begin{proposition}[Conditional independence of trajectories]\label{prop:trajectory}
If $\bm R$ is a ReD-BaSS and the dynamics of $\bm X,\bm Y$ is factorisable,
then $\bm X \independent \bm Y \mid \bm R$. Thus, such system is perfectly
described as a PALO, and $\bm R$ is a MB (in Pearl's sense).
\end{proposition}

A direct consequence of this Proposition is that a ReD-BaSS does not guarantee
statistical independence of $\bm X, \bm Y$ at the trajectory level in
non-factorisable systems. Therefore, in such systems there are interactions
between $\bm X$ and $\bm Y$ that are not mediated by the \MI. Please note that
this is not a weakness of the \MI construction --- which is optimal in
capturing all the directed influences, as shown in Proposition~\ref{lemma:te}.
Instead, this result suggests that non-factorisable systems might not be
well-suited to be described as a PALO.

To further understand this, let us explore the integrated information in the
system $(\bm X,\bm Y)$ using information geometry~\cite{oizumi2016unified}. For
this, consider the manifolds
\begin{align}
\mathcal{M}_1 &= \big\{q_t :
q(x_{t+1},y_{t+1}|\past{x}{t},\past{y}{t})= q(x_{t+1}|\past{x}{t},\past{y}{t})
q(y_{t+1}|\past{x}{t},\past{y}{t})\big\}~, \nonumber \\
\mathcal{M}_2 &= \big\{q_t :
q(x_{t+1},y_{t+1}|\past{x}{t},\past{y}{t})= q(x_{t+1}|\past{x}{t})
q(y_{t+1}|\past{y}{t})\big\}~.  \nonumber
\end{align}
Manifold $\mathcal{M}_1$ corresponds to all systems with factorisable dynamics,
and $\mathcal{M}_2$ to all systems where the dynamics of agent and environment
are fully decoupled. The information-geometric projection of an arbitrary
system $p_t$ onto $\mathcal{M}_2$,
\begin{equation}\label{eq:synergy}
    \tilde{\varphi}_t \coloneqq 
     \min_{q_t\in\mathcal{M}_2} D(p_t|| q_t)~,
\end{equation}
has been proposed as a measure of integrated
information~\cite{ay2015information,mediano2019measuring}. Using the Pythagoras
theorem~\cite{amari2007methods} together with the fact that $\mathcal{M}_2
\subset \mathcal{M}_1$, one can decompose $\tilde\varphi_t$ as
\begin{align}\label{eq:decomp}
  \underbrace{~~~~\tilde{\varphi}_t~~~~\vphantom{\Big[\Big]}}_{D(p_t \| q^{(2)}_t)} =
    \underbrace{~~\Ex\{ \xi_t \}~~\vphantom{\Big[\Big]}}_{D(p_t|| q^{(1)}_t)} +
    \underbrace{\Big[ \text{TE}(\bm A\rightarrow \bm Y)_t + \text{TE}(\bm S\rightarrow \bm X)_t \Big]}_{D(q^{(1)}_t|| q^{(2)}_t)}~,
\end{align}
where $q^{(k)}_t \coloneqq \argmin_{q_t\in\mathcal{M}_k} D(p_t||
q_t)$.\footnote{Note that in non-ergodic scenarios the expected values are not
calculated over individual trajectories, but over the ensemble statistics that
define the probability.}

This decomposition confirms previous results that showed that integrated
information is a construct that combines low-order transfer and high-order
synergies~\cite{mediano2019beyond}. Thanks to Lemma~\ref{lemma:te},
Eq.~\eqref{eq:decomp} states that the transfer component of $\tilde\varphi_t$
(i.e. $D\big(q_t^{(1)}||q_t^{(2)}\big)$) is what is properly mediated by the
\MI. In contrast, the part of $\tilde\varphi$ related to high-order statistics,
i.e. $\Ex\{ \xi_t \} = I(X_{t+1}; Y_{t+1} | \past{X}{t}, \past{Y}{t})$, is not
accounted by the \MI. This last term can either refer to spurious synchronous
correlations (due e.g. to sub-sampling), or be due to synergistic dynamics that
are a signature of emergent phenomena~\cite{rosas2020reconciling}.

In summary, our results suggest that the dynamics of a system $(\bm X,\bm Y)$
that is too synergistically integrated are poorly represented as a PALO, even
if the \MI formally still exists. Additionally, the synergistic component of
integrated information can be used as a measure for this mismatch.

\section{Conclusion}

This manuscript introduced a data-driven method to build PALOs leveraging
principles of computational mechanics. Our construction provides an
informational interpretation of sensory and actuation variables: sensory
(resp.\ active) variables encode all the changes from ``outside'' (resp.\
``inside'') that affect the future evolution of the ``inside'' (resp.
``outside''). Our framework is broadly applicable, depending only on the
underlying bipartition but not imposing any further conditions on the system's
dynamics or distribution. Furthermore, we illustrated how this construction
allows one to relate --- within a PALO framework --- the separation of a system
and its environment to the integrated information encompassing the two.
 
It is to be noted that the CB construction relies on discrete time, which,
while being immediately applicable to digitally sampled data, might not be
natural in some scenarios. Also, CB theory at this stage does not provide
explicit links with probabilistic inference. As shown in Example~\ref{ex:1},
CBs provide a natural extension of Eq.~$\eqref{eq:mbvectorfield}$ to the
discrete-time case, so one possibility would be to combine them with the MB
condition in Eq.~\eqref{eq:mbfactorization}. The exploration of such ``causal
Markov blankets'' which would satisfy both Eq.~\eqref{eq:mbfactorization} and
Definition~\ref{def:MB} is an interesting avenue for future research.

It is our hope that the CB construction may enrich the toolbox of researchers
studying PALOs and help to illuminate further our understanding of the nature
of agency.

\newpage
 
\bibliographystyle{splncs04}
\bibliography{main}

\appendix

\section{Proofs}

\begin{proof}[Lemma~\ref{lemma:te}]
Let's consider $\bm U$ to be a D-BaSS of $\bm X$ w.r.t. $\bm Y$. Then, property
(ii) of a D-Bass is equivalent to
\begin{equation}\label{eq:equiv_cMI}
I(\past{X}{t} ; Y_{t+1} \: | \: U_{t}, \past{Y}{t} ) = 0~.
\end{equation}
Using this, one can verify that
\begin{equation}
I(\past{X}{t} ; Y_{t+1} | \past{Y}{t} ) 
= I(U_{t}, \past{X}{t} ; Y_{t+1} | \past{Y}{t} )
= I(U_{t} ; Y_{t+1} | \past{Y}{t} ) ~. \nonumber
\end{equation}
Here, the first equality holds because $U_{t}$ is a deterministic function of
$\past{X}{t}$, and the second equality follows from an application of the chain
rule and Eq.~\eqref{eq:equiv_cMI}.
\end{proof}

\begin{proof}[Theorem~\ref{prop:unique}]
Consider the function $F(\cdot)$ that maps each $\past{x}{t}$ to its
corresponding equivalence class $F(\past{x}{t})$ established by the equivalence
relationship $\equiv_p$, and define $M_t = F(\past{X}{t})$. As this
construction satisfies the requirement of precedence in Def.~\ref{def:DBass},
let us show the sufficiency of $\bm M$. By definition of $M_t$, it is clear
that if $m_t = F(\past{x}{t})$ then
\begin{equation}
p(y_{t+1}| \past{x}{t}, \past{y}{t}) = p(y_{t+1}| m_{t}, \past{y}{t}) ~, \nonumber
\end{equation}
which implies that $H(Y_{t+1} | \past{X}{t}, \past{Y}{t} ) = H(Y_{t+1} | M_{t},
\past{Y}{t} )$. As a consequence,
\begin{align}
I(\past{X}{t} ; Y_{t+1} | \past{Y}{t} ) &= H(Y_{t+1} | \past{Y}{t} ) - H(Y_{t+1} | \past{X}{t}, \past{Y}{t} ) \nonumber\\
&= H(Y_{t+1} | \past{Y}{t} ) - H(Y_{t+1} | M_{t}, \past{Y}{t} ) \nonumber \\
&= I( M_{t} ; Y_{t+1} | \past{Y}{t} ) ~.\label{eq:cool}
\end{align}
From this, sufficiency follows from noticing that
\begin{align}
I(\past{X}{t} ; Y_{t+1} | M_t, \past{Y}{t} ) &= I(\past{X}{t}, M_t ; Y_{t+1} | \past{Y}{t} ) - I(M_t ; Y_{t+1} | \past{Y}{t} ) \nonumber\\
&= I(\past{X}{t} ; Y_{t+1} | \past{Y}{t} ) - I(M_t ; Y_{t+1} | \past{Y}{t} ) \nonumber\\
&= 0~. \nonumber
\end{align}
Above, the first equality is due to the chain rule, the second follows from the
fact that $M_t$ is a function of $\past{X}{t}$, and the third uses
Eq.~\eqref{eq:cool}.

To finish the proof, let us show that $\bm M$ is minimal. For this, consider
another $\bm U$ to be another D-BaSS of $\bm X$ w.r.t. $\bm Y$. As $U_t =
G(\past{X}{t})$ for some function $G(\cdot)$, $\bm U$ corresponds to another
partition of the trajectories $\past{x}{t}$. If there exists no function $f$
such that $f(U_t) = M_t$, that implies that the partition that corresponds to
$\bm M$ is not a coarsening of the partition for $\bm U$, and therefore that
there exists $\past{x}{t}$ and $\past{x}{t}'$ such that $G(\past{x}{t}) =
G(\past{x}{t}')$ while $p(y_{t+1}| \past{x}{t}, \past{y}{t}) \neq p(y_{t+1}|
\past{x}{t}', \past{y}{t})$. This, in turn, implies that there exists a
$\past{x}{t}'$ such that that $p(y_{t+1}| u_t,\past{x}{t}', \past{y}{t}) \neq
p(y_{t+1}| u_t, \past{y}{t}) = \sum_{\past{x}{t}} p(y_{t+1}| u_t,\past{x}{t},
\past{y}{t}) p(\past{x}{t} | u_t,\past{y}{t})$, showing that $\past{X}{t}$ is
not conditionally independent of $Y_{t+1}$ given $U_t, \past{Y}{t}$,
contradicting the fact that $\bm U$ is a D-BaSS. This contradiction proves that
the partition induced by $\bm U$ is a refinement of the partition induced by
$\bm M$, proving the minimality of the latter.
\end{proof}

\begin{proof}[Proposition~\ref{prop:PALO}]
Let's denote by $\bm A$ the minimal D-BaSS of $\bm X$ w.r.t. $\bm Y$, and $\bm
S$ the minimal D-BaSS of $\bm Y$ w.r.t. $\bm X$, which are known to exist and
be unique thanks to Theorem~\ref{prop:unique}. Then, by defining $M_t \coloneqq
(S_t,A_t)$, one can directly verify that $\bm M$ is a ReD-BaSS of $(\bm X,\bm
Y)$. To prove its minimality, let us consider another ReD-BaSS of $(\bm X,\bm
Y)$ denoted by $\bm N$. As $\bm N$ is a D-BaSS of $\bm X$ w.r.t. $\bm Y$, the
minimality of $\bm A$ guarantees the existance of a mapping $f(\cdot)$ such
that $f(N_t) = S_t$. Similarly, thanks to the minimality of $\bm S$, there is
another mapping $g(\cdot)$ such that $g(N_t) = A_t$. Therefore, the function
$F(\cdot) = (f,g)$ satisfies $F(N_t) = M_t$, which confirms the minimality of
$\bm M$.
\end{proof}

\begin{proof}[Proposition~\ref{prop:trajectory}]
The proof is based on the principle that if $p(A,B,C) = f(A,C) g(B,C)$ , then
$A\independent B | C$. Building on that rationale, a direct calculation shows
that
\begin{align}
p(\bm x, \bm y) &= \prod_{\tau=-\infty}^{\infty} p(x_{\tau+1}, y_{\tau+1} | \past{x}{\tau},  \past{y}{\tau} ) \nonumber \\
& = \prod_{\tau=-\infty}^{\infty} \exp\{\xi_\tau\} \: p(x_{\tau+1} | \past{x}{\tau},  \past{y}{\tau} ) \: p(y_{\tau+1} | \past{x}{\tau},  \past{y}{\tau} ), \label{eq:decnonfact}
\end{align}
where the second equality\footnote{Note that the infinite products in this
proof are just a formal procedure to acknowledge products that can be taken up
to arbitrary times.} uses Eq.~\eqref{def:syn}. Additionally, if, as per
assumption of the Proposition, $\bm R$ is a ReD-BaSS of $(\bm X,\bm Y)$, then
\begin{equation}
p(x_{\tau+1} | \past{x}{\tau}, \past{y}{\tau} ) = p(x_{\tau+1} | \past{x}{\tau}, \past{y}{\tau} , r_{\tau} ) 
= p(x_{\tau+1} | \past{x}{\tau}, r_{\tau}  ), \nonumber
\end{equation}
where the first equality uses the fact that $r_\tau$ (by definition) is a
function of $(\past{x}{\tau}, \past{y}{\tau})$, and the second uses the
sufficiency of D-BaSS's. Following an analogous derivation, one can show that
$p(y_{\tau+1} | \past{x}{\tau}, \past{y}{\tau} ) = p(y_{\tau+1} | r_{\tau},
\past{y}{\tau} )$. Then, with the assumption that the dynamics of $(\bm X,\bm
Y)$ is factorisable and hence $\xi_t=0$, it follows from
Eq.~\eqref{eq:decnonfact} that
\begin{equation}
p(\bm x, \bm y) 
= \prod_{\tau=-\infty}^{\infty} p(x_{\tau+1} | r_{\tau}, \past{y}{\tau} ) \: p(y_{\tau+1} | r_{\tau}, \past{y}{\tau} )~.
\nonumber
\end{equation}
Separating the two product series, this shows that there exist functions
$f(\cdot)$ and $g(\cdot)$ such that $p(\bm x, \bm y) = f(\bm x, \bm r) g(\bm y,
\bm r)$, and hence one has $\bm X \independent \bm Y | \bm R~$, which completes
the proof.
\end{proof}

\end{document}